\documentclass[a4paper,10pt]{article}

\newcommand{\sect}[1]{\setcounter{equation}{0}\section{#1}}

\begin{document}
\topmargin 0pt \oddsidemargin 0mm

\newcommand{\be}{\begin{equation}}
\newcommand{\ee}{\end{equation}}
\newcommand{\tg}{\mbox{}\indent}
\def\disp{\displaystyle}

\renewcommand{\thefootnote}{\fnsymbol{footnote}}
\begin{titlepage}
\vspace{4cm}
\begin{flushright}
 hep-th/0409130
\end{flushright}

\vspace{5mm}
\begin{center}
{\Large \bf Second-Order Corrections to the Power Spectrum in the
Slow-Roll Expansion with a Time-Dependent Sound Speed}
\vspace{32mm}

{\large Hao Wei~$^{1,2}$\footnote{e-mail address:
haowei@itp.ac.cn}, Rong-Gen Cai~$^{1,3}$\footnote{e-mail address:
cairg@itp.ac.cn} and Anzhong Wang~$^3$\footnote{e-mail address:
anzhong\_wang@baylor.edu}}

\vspace{10mm}
 {\em $~^1$ Institute of Theoretical Physics, Chinese
Academy of Sciences,\\
 P.O. Box 2735, Beijing 100080, China \\
  $^2$ The Graduate School of the Chinese Academy of Sciences \\
 $~^3$  CASPER, Department of Physics, Baylor University, Waco,
 TX76798-7316, USA}

\end{center}

\vspace*{35mm}
\centerline{{\bf{Abstract}}}
 \vspace{15mm}
We extend Green's function method developed by Stewart and Gong to
calculate the power spectrum of density perturbation in the case
with a time-dependent sound speed, and explicitly give the power
spectrum and spectral index up to second-order corrections in the
slow-roll expansion. The case of tachyon inflation is included as
a special case.
\end{titlepage}

\newpage

\vskip 7mm

\sect{Introduction}

 It is generally believed that the curvature perturbation produced
during inflation \cite{s1} is the origin of the inhomogeneities
necessary for large scale structure formation. The power spectrum
of these perturbations is predicted to be approximately scale
invariant by most of the inflation scenarios \cite{s2}, and
 this is confirmed by many recent cosmic microwave background observations
 and galaxy surveys \cite{s3}. The observation results from WMAP \cite{s4},
  SDSS \cite{s5} and other experiments give
 more accurate measurement of the power spectrum and the spectral index than ever before.
  Thus, it is very important to calculate the power spectrum precisely to match the
  observation results fully and to distinguish different inflation models.

 Many works have been done on the power spectrum for the density perturbations
produced during inflation~\cite{s6}. For example, the power
spectrum and spectral index of density fluctuation are calculated
up to the first order of slow-roll parameters in~\cite{insert1}.
In \cite{s7} Stewart and Gong further set up a formalism that can
be used to calculate the power spectrum of the curvature
perturbations  up to arbitrary order in the slow-roll expansion,
and calculate the power spectrum and spectral index up to
second-order corrections.  Recently, Myung et al.\cite{s8} have
calculated the second-order corrections in the brane inflation
model and in the noncommutative space-time inflation model,
respectively.

However, we note that in most calculations made before, the sound
speed $C_{s}^{2}$ is assumed to be constant. Of cause, it is true
in the inflation models derived by a scalar field of canonical
kinetic term, where $C_{s}^{2}=1$. But, as we will see in the
following, the sound speed is time-dependent in the general case.
For instance, in the case of tachyon inflation (see for
example~\cite{s11,s12} and references therein), which has received
a lot of attention in recent years,
 $C_{s}^{2}=1-\frac{2}{3}\epsilon$, where $\epsilon\equiv -\dot{H}/{H^2}$ is
 a first-order slow-roll parameter and $H$ is the Hubble parameter, a dot denotes
  the derivative with respect to cosmic time $t$. If one only considers
  approximations up to
  the first-order corrections, because $\dot{\epsilon}$ is a second-order
small parameter, $\epsilon$ and then $C_{s}^{2}$ can be treated as
constant approximately. But when one attempts to calculate
second-order corrections, one has to consider the time-dependence
of these quantities.  As a result, the Mukhanov equation governing
 quantum fluctuations produced during inflation becomes more difficult to solve than the
 case with a constant sound speed.

In this paper, we have extended Green's function method developed
by Stewart and Gong \cite{s7} and obtained the second-order
corrections in the slow-roll
 expansion to the power spectrum in the case with a  time-dependent
 sound speed. In particular, the case of tachyon inflation is included.
Note that the formalism can be used to calculate the power
spectrum of the curvature perturbation up to arbitrary order in
the slow-roll expansion. We use the units $\hbar=c=8\pi G=1$
throughout this paper.

\sect{Preliminaries}

\subsection{Basics}
 Our starting point is the following effective action during inflation~\cite{s9},
\be \label{eq1}
 S=\int d^{4}x\sqrt{-g}\left[-\frac{1}{2}R+p(\varphi, X)\right] \ee where
$g$ is the determinant of the metric, $R$ is the Ricci scalar,
$\varphi$ denotes a scalar field and
 \be
  \label{eq2}
X=\frac{1}{2}g^{\mu\nu}\partial_{\mu}\varphi\partial_{\nu}\varphi,
\ee
 is its kinetic term. The Lagrangian $p$ plays the role of pressure and the energy
density is given by
\be
 \label{eq3}
 \rho=2Xp_{,_X}-p.
\ee
 Here $p_{,_X}$ denotes the partial derivative of $p$ with
respect to $X$. In this paper, we consider a flat universe for
simplicity, namely $K=0$. The background metric is
 \be
 \label{eq4}
ds^2=dt^2-a^{2}(t){\bf dx}^2=a^{2}(\eta)\left(d\eta^2-{\bf
dx}^2\right),
  \ee
where $\eta$ is the conformal time defined by $d\eta=dt/a$.  The
action for scalar perturbation can be written down as~\cite{s9}
 \be
 \label{eq5}
S=\frac{1}{2}\int\left[w^{\prime\,2}+C_{s}^{2}w\nabla^{2}
w+\frac{z^{\prime\prime}}{z}w^2\right]d\eta d^{3}x,
\ee
 where a prime
denotes the derivative with respect to the conformal time $\eta$,
the canonical quantization variable $w=z\zeta$, $\zeta$ is the
curvature perturbation, and $z$ is defined by
 \be
 \label{eq6}
z\equiv\frac{a(\rho+p)^{1/2}}{C_{s}H}, \ee
 and the sound speed by
 \be
 \label{eq7}
  C_{s}^{2}\equiv\frac{p_{,_X}}{\rho_{,_X}}.
  \ee
Using the Fourier transform of $w$, from (\ref{eq5}) one can
obtain the equation of motion for mode $w_k$, i.e. the well-known
Mukhanov equation~\cite{s9}
 \be
 \label{eq8}
w_{k}^{\prime\prime}+\left(C_{s}^{2}k^2-\frac{z^{\prime\prime}}{z}\right)w_{k}=0,
\ee where $k$ is the wave-number for the mode $w_k$. Furthermore,
from Eq.~(\ref{eq6}) and the dynamical equation of the Hubble
parameter \be
  \label{eq9}
\dot{H}=-\frac{1}{2}(\rho+p), \ee we can express $z$ by the
slow-roll parameter $\epsilon\equiv -\dot{H}/H^2$, which is a
first-order slow-roll parameter, as
 \be
 \label{eq10}
z=\sqrt{2}\,{a\over{C_s}}\,\epsilon^{1/2}.
 \ee

\subsection{Sound speed}

 Obviously, the sound speed $C_s$ defined by Eq.~(\ref{eq7}) is
 generally
time-dependant for a general Lagrangian $p(\varphi,X)$. In the
slow-roll phase where $X$ is a small quantity, compared to the
potential of the scalar field $\varphi$. As a result, a generic
Lagrangian is expected to admit an expansion of the
form~\cite{s10}
 \be
 \label{eq11}
p(\varphi,X)=V(\varphi)+D(\varphi)X+B(\varphi)X^2+\cdots.
\ee
 From
Eqs. (\ref{eq11}), (\ref{eq3}) and (\ref{eq7}), one has the sound
speed
 \be
 \label{eq12}
C_{s}^{2}=1-\frac{4BX}{D}+24\left(\frac{BX}{D}\right)^2,
 \ee
 up to the term of $X$ squared. Substituting Eqs.~(\ref{eq11}) and (\ref{eq3}) into
 Eq.~(\ref{eq9}), we
get \be \label{eq13}
 2BX^2+DX-H^{2}\epsilon=0.
  \ee
  This can be considered as a quadratic equation of $X$ and the solution
satisfying the convergence condition $DX\gg BX^2$ is
 \be
 \label{eq14}
X=\frac{H^2}{D}\epsilon-{2BH^4\over D^2}\epsilon^{2},
 \ee
 up to the second-order correction of slow-roll parameter.
  Then, one can expand the sound speed in terms of the
  slow-roll parameter as
  \be
  \label{eq15}
C_{s}^{2}=1-\frac{4BH^2}{D^2}\epsilon+\frac{8B{^2}H^{4}}{D^3}
   \left(1+\frac{3}{D}\right)\epsilon^2.
\ee
 Because the coefficients $B$ and $D$ are dependent of
$\varphi$, and in turn dependent of time, it is still very
difficult to deal with Eq.~(\ref{eq8}). From now on, we restrict
ourselves to a class of special case where the sound speed has the
form
 \be
 \label{eq16}
  C_{s}^{2}=1-c_{1}\epsilon-c_{2}\epsilon^{2}.
   \ee
Here $c_1$ and $c_2$ are two constants of order of unity. This
form of sound speed includes some well-known examples. For
instance,
 in the case of  canonical scalar field  one has $c_1=c_2=0$
 while $c_1=2/3$, $c_2=0$ for the case of tachyon
 inflation \cite{s11,s12}.

 In fact, by using the formalism developed in section 3, we can consider a class
 of more general sound speed as
$$C_{s}^{2}=1-c_{1}\epsilon_{m}-c_{2}\epsilon_{p}\epsilon_{q},$$
where $\epsilon_{i}$ is defined by Eq. (\ref{eq17}) and $i=m$,
$p$, $q$ are some positive integers. We will give a simple
discussion on this in the appendix. But here, we will concentrate
ourselves on the case with the sound speed given by
Eq.~(\ref{eq16}).

\subsection{Slow-roll parameters}

 In this paper, we introduce a  set of slow-roll parameters
 expressed by Hubble parameter and its derivatives with respect to time
 as
\be
 \label{eq17}
\epsilon_{n}\equiv\frac{(-1)^n}{H}\frac{d^{n}H}{dt^{n}}\left/\frac{d^{n-1}H}{dt^{n-1}}\right.,
\ee namely, one has \be \label{eq18}
 \epsilon_1=\epsilon\equiv
-\frac{\dot{H}}{H^2},~~~~~~~~\epsilon_2\equiv\frac{\ddot{H}}{H\dot{H}},
~~~~~~~\epsilon_3\equiv
-\frac{\stackrel{\dots}{H}}{H\ddot{H}},\hspace{3mm}\dots \ee
evaluated at sound horizon crossing~\cite{s9}, i.e. $aH=C_sk$.
Note that, different from other slow-roll parameters used for
example in \cite{s7,s2}, these parameters defined in (\ref{eq17})
are all first-order small parameters, namely satisfying
$|\epsilon_{n}|<\xi$ for some small perturbation parameter $\xi$.
The higher order slow-roll parameters are products of the
first-order parameters $\epsilon_n$.  In addition, it is easy to
show
 \be
 \label{eq19}
\frac{1}{H}\,\dot{\epsilon}_n=\epsilon_{n}\left[\epsilon_1+(-1)^{n+1}
(\epsilon_n+\epsilon_{n+1})\right],
 \ee
which will be used extensively below.

Note that for the case of canonical scalar field, the slow-roll
parameters used in Ref.~\cite{s7} are
$$\epsilon_1\equiv -\frac{\dot{H}}{H^2}=\frac{1}{2}\left(\frac{\dot{\phi}}{H}\right)^{2},
~~~~~~~
\delta_{n}\equiv\frac{1}{H^{n}\dot{\phi}}\frac{d^{n+1}\phi}{dt^{n+1}},$$
and these have the following relations to the slow-roll parameters
defined in (\ref{eq17})
 \be
 \label{eq20}
\epsilon_1=\epsilon_1,
 ~~~~~\delta_1=\frac{1}{2}\epsilon_2,
 ~~~~~\delta_2=-\frac{1}{4}\epsilon_{2}^{2}-\frac{1}{2}\epsilon_{2}\epsilon_{3},
 ~~~~~\delta_3=\frac{3}{8}\epsilon_{2}^{3}+\frac{3}{4}\epsilon_{2}^{2}\epsilon_{3}
-\frac{1}{2}\epsilon_{2}\epsilon_{3}\epsilon_{4},\ \ \dots
 \ee

\sect{The calculations}

\subsection{Slow-roll expansion and the formalism of calculations}

 Our main task is to solve the Mukhanov equation
Eq.~(\ref{eq8}). Introducing new variables $y=\sqrt{2k}w_{k}$ and
$x=-k\eta$, we can recast the Mukhanov equation as
\be
 \label{eq21}
\frac{d^{2}y}{dx^2}+\left(C_{s}^{2}-\frac{1}{z}\frac{d^{2}z}{dx^2}\right)y=0.
\ee
Because $C_{s}^{2}$ is time-dependent, the problem is more
complicated than the case with constant $C_s$. In order to obtain
the solution of (\ref{eq21}) up to second order corrections in the
slow-roll expansion, we develop a formalism, in which the key
point is to use Green's function method developed in
Ref.~\cite{s7} twice. The first step is to decompose the term
$\disp\frac{1}{z}\frac{d^{2}z}{dx^2}=\frac{1}{k^2}\frac{z^{\prime\prime}}{z}$
into two parts and Eq.~(\ref{eq21}) then can be rewritten as
 \be
  \label{eq22}
\frac{d^{2}y}{dx^2}+\left(C_{s}^{2}-\frac{2}{x^2}\right)y
 =\frac{1}{x^2}g(\ln x)y,
\ee
where $g(\ln x)$ is defined by
\be
 \label{eq23}
  g(\ln x)=\frac{x^2}{k^2}\frac{z^{\prime\prime}}{z}-2.
  \ee
  We can justify
this decomposition as follows. From Eq.~(\ref{eq10}) we have
\be
\frac{z^{\prime\prime}}{z}=\frac{a^{\prime\prime}}{a}
+\left(\frac{a^{\prime}}{a}\right)\left(\frac{\epsilon_{1}^{\prime}}{\epsilon_1}\right)
-2\left(\frac{a^{\prime}}{a}\right)\left(\frac{C_{s}^{\prime}}{C_{s}}\right)
-\frac{1}{4}\left(\frac{\epsilon_{1}^{\prime}}{\epsilon_1}\right)^2
+\frac{1}{2}\left(\frac{\epsilon_{1}^{\prime\prime}}{\epsilon_1}\right)
-\left(\frac{\epsilon_{1}^{\prime}}{\epsilon_1}\right)\left(\frac{C_{s}^{\prime}}{C_{s}}\right)
-\frac{C_{s}^{\prime\prime}}{C_s}+2\left(\frac{C_{s}^{\prime}}{C_{s}}\right)^2.
\ee
Note that
 \be
 \label{eq25}
\frac{a^{\prime\prime}}{a}=a^{2}H^{2}(2-\epsilon_{1}),
 \ee
 and
 \be
 \label{eq26}
x=-k\eta=-k\int\frac{dt}{a}=-k\int\frac{da}{a^{2}H}
 \simeq\frac{k}{aH}\left(1+\epsilon_{1}+3\epsilon_{1}^{2}
 +\epsilon_{1}\epsilon_{2}\right).
\ee Thus, for any $C_{s}^{2}$, the leading term of
$\disp\frac{1}{z}\frac{d^{2}z}{dx^2}$ is $\disp\frac{2}{x^2}$,
which comes from $\disp\frac{a^{\prime\prime}}{a}$. Therefore, the
decomposition made above is always valid. Furthermore, we can
expand
 \be \label{eq27}
  g(\ln x)=\sum\limits_{n=0}^{\infty}\frac{g_{n+1}}{n!}
  \left(\ln x\right)^{n},
\ee where $g_n$ is of order $n$ in the slow-roll expansion, namely
$|g_{n}|<\xi^{n}$. This expansion is useful for $\exp(-1/\xi)\ll
x\ll \exp(1/\xi)$. From Eqs.~(\ref{eq23})-(\ref{eq27}),
(\ref{eq16}) and (\ref{eq19}), up to second-order corrections, it
is easy to find
 \be
 \label{eq28}
 g_2=\left.\frac{dg}{d\ln x}\right|_{x=1}\simeq \left.\left(-12\epsilon_{1}^{2}
 -\frac{15}{2}\epsilon_{1}\epsilon_{2}
+\frac{3}{2}\epsilon_{2}^{2}+\frac{3}{2}\epsilon_{2}\epsilon_{3}\right)\right|_{aH=C_sk},
\ee
 \be
 \label{eq29}
  g_{1}=g|_{x=1}\simeq  \left.\left[\,
6\epsilon_{1}+(20+3c_{1})\epsilon_{1}^{2}+\frac{3}{2}\epsilon_{2}+(9+\frac{3}{2}c_{1})
\epsilon_{1}\epsilon_{2}-\frac{1}{4}\epsilon_{2}^{2}-\frac{1}{2}\epsilon_{2}
\epsilon_{3}\right]\right|_{aH=C_sk}. \ee Note that  here $\ln x\,
|_{aH=C_sk}$ is a first-order small quantity in the slow-roll
expansion.

 The boundary condition of Eq.~(\ref{eq21}) or Eq.~(\ref{eq22}) required to get the
 power spectrum is
 \be
 \label{eq30}
 y\to\left\{
\begin{array}{ll}\lim\limits_{x\to\infty}{\bar{y}}_{0}(x) & {\rm
as}\hspace{2mm}x\to\infty,\\ \\ \sqrt{2k}A_{k}z & {\rm
as}\hspace{2mm}x\to 0,
\end{array}\right.
 \ee
 where $\bar{y}_{0}(x)$ is a solution of equation
  \be
  \label{eq31}
\frac{d^{2}y}{dx^2}+\left(C_{s}^{2}-\frac{2}{x^2}\right)y=0.
 \ee
Note that $C_{s}^{2}$ is time-dependent, and then $x$-dependant.
To solve Eq.~(\ref{eq31}), it is necessary to express $C_{s}^{2}$
as a function of $x$. Since $\epsilon_{1}=-\dot{H}/H^2$ is
time-dependant and therefore is a function of $x$, too. From
Eqs.~(\ref{eq19}) and (\ref{eq26}), up to second-order, we have
$$\frac{d\epsilon_1}{dx}\simeq -x^{-1}\epsilon_{1}(2\epsilon_{1}+\epsilon_{2}).$$
 Here $\epsilon_{1}(2\epsilon_{1}+\epsilon_{2})$ can be treated as a
constant approximately because its derivative with respect to time
is a third-order small quantity, which can be ignored. Therefore
$\epsilon_1$ should be of the form of $const.+const.\ln x$, up to
second-order corrections. In fact, we can expand $\epsilon_{1}$ as
\be
 \label{eq32}
 \epsilon_{1}=\sum\limits_{n=0}^{\infty}\frac{s_{n+1}}{n!}(\ln x)^{n},
  \ee
   where $s_n$ is of order $n$ in the slow-roll
expansion, namely $|s_{n}|<\xi^{n}$. This expansion is useful for
$\exp(-1/\xi)\ll x\ll \exp(1/\xi)$. It is then easy to obtain
 \be
 \label{eq33}
 s_2=\left.\frac{d\epsilon_1}{d\ln x}\right|_{x=1}\simeq
 \left.-\epsilon_{1}(2\epsilon_{1}+\epsilon_{2})\right|_{aH=C_sk},
\ee
 \be
 \label{eq34}
s_1=\left.\epsilon_{1}\right|_{x=1}\simeq\left.\epsilon_{1}\right|_{aH=C_sk}.
\ee
 Substituting Eqs.~(\ref{eq32})-(\ref{eq34}) into Eq.~(\ref{eq16}),
 up to
second-order, we can recast the sound speed as
 \be
  \label{eq35}
C_{s}^{2}=C_{sc}^{2}-c_{1}s_{2}\ln x,
\ee
in which
\be
 \label{eq36}
C_{sc}^{2}=1-c_{1}s_{1}-c_{2}s_{1}^{2}.
 \ee
 Then, Eq.~(\ref{eq31}) can be approximately rewritten as
  \be
  \label{eq37}
\frac{d^{2}y}{dx^2}+\left(C_{sc}^{2}-\frac{2}{x^2}\right)y
 =(c_{1}s_{2}\ln x)y.
  \ee
  Using  Green's function approach, Equation (\ref{eq37}) can be
  perturbatively solved order by order. The solution of Eq.~(\ref{eq31}) or (\ref{eq37})
   is
   \be
   \label{eq38}
 \bar{y}_{0}(x)=y_{0}(x)+\frac{i\,c_{1}}{2C_{sc}}\int_{x}^{1}du\;s_{2}
 \ln u\;\bar y_0(u)[y_{0}^{*}(u)y_{0}(x)-y_{0}^{*}(x)y_{0}(u)],
\ee
where
 \be
 \label{eq39}
y_{0}(x)=\left(1+\frac{i}{C_{sc}x}\right)e^{i\, C_{sc}\, x}
\ee
satisfying
 \be
 \label{eq40}
\frac{d^{2}y_{0}}{dx^2}+\left(C_{sc}^{2}-\frac{2}{x^2}\right)y_{0}=0.
\ee
 Ignoring the 4th- and higher order corrections, from (\ref{eq38}) we get
\be
 \label{eq41}
 \bar{y}_{0}(x)=y_{0}(x)+\frac{i\,c_{1}}{2C_{sc}}
  \int_{x}^{1}du\;s_{2}\ln u\;y_{0}(u)[y_{0}^{*}(u)y_{0}(x)
  -y_{0}^{*}(x)y_{0}(u)]\equiv y_{0}(x)+y_{s}(x).
 \ee
 Now, we are able to solve Eq.~(\ref{eq22})
by using Green's function approach once again. The solution is
\begin{eqnarray}
 \label{eq42}
y(x)&\simeq& \bar{y}_{0}(x)+\disp\frac{i}{2C_{sc}}
\int_{x}^{\infty}du\frac{1}{u^2}g(\ln
u)y(u)[\bar{y}_{0}^{*}(u)\bar{y}_{0}(x)
-\bar{y}_{0}^{*}(x)\bar{y}_{0}(u)] \nonumber
\\
&\simeq&\bar{y}_{0}(x)+\disp\frac{i}{2C_{sc}}
\int_{x}^{\infty}du\frac{1}{u^2}g(\ln
u)y(u)[y_{0}^{*}(u)y_{0}(x)-y_{0}^{*}(x)y_{0}(u)],
 \end{eqnarray}
  in which we
have ignored the third and higher order corrections.  Using
Eq.~(\ref{eq42}) twice, we have the perturbative solution up to
second-order corrections
 \be
 \label{eq43}
y(x)=y_{0}(x)+y_{s}(x)+y_{1}(x)+y_{21}(x)+y_{22}(x),
 \ee
  where
$y_{s}(x)$ comes from Eq.~(\ref{eq41}), which is a second-order
correction to $y(x)$, and given by
\begin{eqnarray}
y_{s}(x)&=&\disp\frac{i\, c_{1}}{2C_{sc}} \disp\int_{x}^{1}du\,
      s_{2}\ln u\, y_{0}(u)[y_{0}^{*}(u)y_{0}(x)-y_{0}^{*}(x)y_{0}(u)] \nonumber \\
&=& \disp\frac{i\,c_{1}s_{2}}{2C_{sc}}\left\{y_{0}(x)\left[-1-x\ln
    x+x+ \frac{1}{C_{sc}^{2}}(x^{-1}\ln x-1+x^{-1})\right]\right.
    -y_{0}^{*}(x) \left[\disp\frac{i}{2C_{sc}}\times\right. \nonumber \\
& & \left.\left. e^{2i\,C_{sc}\,x}\ln x
   -\disp\frac{3\,i}{2C_{sc}}\int_{x}^{1}du\,u^{-1}e^{2iC_{sc}u}-\frac{1}{C_{sc}^{2}}
   \left(x^{-1}e^{2iC_{sc}x}(\ln
   x+1)-e^{2iC_{sc}}\right)\right]\right\}.
\end{eqnarray}
 $y_{1}(x)$ is a first-order correction to $y(x)$,
 \begin{eqnarray}
 \label{eq45}
y_{1}(x)&=&\disp\frac{i\, g_{1}}{2C_{sc}}
   \int_{x}^{\infty}du\, u^{-2}y_{0}(u)[y_{0}^{*}(u)y_{0}(x)-y_{0}^{*}(x)y_{0}(u)]\nonumber \\
&=&-\disp\frac{1}{3}g_{1}\left[-\frac{2\, i}{C_{sc}}x^{-1}e^{iC_{sc}x}+y_{0}^{*}(x)
\int_{x}^{\infty}du\, u^{-1}e^{2iC_{sc}u}\right],
\end{eqnarray}
and the second-order corrections $y_{21}(x)$ and $y_{22}(x)$ are,
respectively, given by
 \begin{eqnarray}
 \label{eq46}
y_{21}(x)&=&\disp\frac{i\, g_{1}}{2C_{sc}} \int_{x}^{\infty}du\,
u^{-2}y_{1}(u)[y_{0}^{*}(u)y_{0}(x)-y_{0}^{*}(x)y_{0}(u)]
 \nonumber \\
 &=&-\disp\frac{i\,
g_{1}^{2}}{9}\left[\,\frac{2}{3C_{sc}}x^{-1}e^{iC_{sc}x}
+\left(-\frac{5}{3C_{sc}}x^{-1}+\frac{i}{3}\right)
\int_{x}^{\infty}du\, u^{-1}e^{2iC_{sc}u}\right. \nonumber \\
& &\left. +i\, y_{0}(x)\disp\int_{x}^{\infty}du\,
u^{-1}e^{-2iC_{sc}u} \int_{u}^{\infty}dv\,
v^{-1}e^{2iC_{sc}v}\,\right],
\end{eqnarray}
and
 \begin{eqnarray}
 \label{eq47}
y_{22}(x)&=&\disp\frac{i\, g_{2}}{2C_{sc}}
 \int_{x}^{\infty}du\, u^{-2}\ln u\, y_{0}(u)[y_{0}^{*}(u)y_{0}(x)-y_{0}^{*}(x)y_{0}(u)]
  \nonumber \\
&=&\disp\frac{i\, g_{2}}{3}\left[\,\frac{1}{C_{sc}}x^{-1}
\left(\frac{8}{3}+2\ln x\right)e^{iC_{sc}x}+\frac{7\,
i}{3}y_{0}^{*}(x) \int_{x}^{\infty}du\, u^{-1}e^{2iC_{sc}u}\right.
 \nonumber \\
& & \left. +i\, y_{0}^{*}(x)\disp\int_{x}^{\infty}du\, u^{-1}\ln
u\, e^{2iC_{sc}u}\,\right].
\end{eqnarray}\
 Note that in writing $y_{22}(x)$ we have ignored a term
$$\disp\frac{i\, g_{2}\, y_{0}(x)}{6\, C_{sc}^{2}}\, x^{-3}
\left(1-\frac{1}{C_{sc}}\right)\left(\ln x+\frac{1}{3}\right),$$
since it is a third-order correction. In addition, the
perturbative solution Eq.~(\ref{eq43}) should be accurate for
$\exp(1/\xi)\gg x\gg \exp(-1/\xi)$.

\subsection{The power spectrum}

Having obtained the solution $y(x)$ of the Mukhanov equation
(\ref{eq22}), we are in a position to calculate the power spectrum
of density fluctuation during inflation.  Taking the limit $x \to
0$ while keeping $\xi\ln (1/x)$ fixed and small, after some
tedious calculations, we obtain the asymptotic forms for
$y_{0}(x)$, $y_{s}(x)$, $y_{1}(x)$, $y_{21}(x)$ and $y_{22}(x)$.
They are
 \be
 \label{eq48}
y_{0}(x) \to \frac{i}{C_{sc}}x^{-1},
 \ee
 \be
 \label{eq49}
  y_{s}(x)\to\frac{3\,i\, c_{1}s_{2}}{4\, C_{sc}^{3}}\,
   x^{-1}\left[\,\frac{4}{3}-\frac{2\, i}{3\, C_{sc}}-\frac{2\, i\,
   C_{sc}}{3}+\frac{2\, i\, e^{2iC_{sc}}}{3\,
   C_{sc}}+\alpha-2+Ei(2iC_{sc})-\frac{i\pi}{2}-\ln
   C_{sc}\,\right],
 \ee
 \be
 \label{eq50}
 y_{1}(x)\to \frac{i\, g_{1}}{3\, C_{sc}}\,
x^{-1}\left[\,\alpha+\frac{i\,\pi}{2}-\ln
C_{sc}\,\right]-\frac{i\, g_{1}}{3\, C_{sc}}\, x^{-1}\ln x,
 \ee
\begin{eqnarray}
\label{eq51}
 y_{21}(x) &\to & \disp\frac{i\, g_{1}^{2}}{18\,
C_{sc}}\left[\,\alpha^{2}-\frac{2}{3}
\alpha-4+\frac{\pi^2}{4}+(\ln
C_{sc})^{2}+\left(\frac{2}{3}-2\alpha\right)\ln C_{sc}
+i\pi\left(\alpha-\frac{1}{3}-\ln C_{sc}\right)\right]x^{-1}
\nonumber \\
 & &-  \disp\frac{i\, g_{1}^{2}}{9\,
C_{sc}}\left[\,\alpha-\frac{1}{3}-\ln C_{sc}
+\frac{i\pi}{2}\,\right]x^{-1}\ln x+\frac{i\, g_{1}^{2}}{18\,
C_{sc}}\, x^{-1}(\ln x)^{2},
\end{eqnarray}
and
\begin{eqnarray}
\label{eq52}
 y_{22}(x) &\to & \disp\frac{i\, g_{2}}{6\,
C_{sc}}\left[\,\alpha^{2}+\frac{2}{3}
 \alpha-\frac{\pi^2}{12}+i\pi\left(\alpha+\frac{1}{3}-\ln C_{sc}\right)-
  2\left(\alpha+\frac{1}{3}\right)\ln C_{sc}+(\ln C_{sc})^{2}\right]x^{-1} \nonumber \\
& &- \disp\frac{i\, g_{2}}{9\, C_{sc}}\, x^{-1}\ln x-\frac{i\,
g_{2}}{6\, C_{sc}}\, x^{-1}(\ln x)^{2},
\end{eqnarray}
respectively, where $\alpha\equiv 2-\ln 2-\gamma$, $\gamma\simeq
0.5772156649$ is the Euler-Mascheroni constant, and
$Ei(z)\equiv\disp\int_{-\infty}^{z}\frac{e^{\chi}}{\chi}d\chi$
is the exponential integral function. Especially, $Ei(2i)\simeq 0.4229808288+i\, 3.1762093036$.\\

 The exact asymptotic form of $y(x)$ in the limit $x\to 0$ is given by Eq.~(\ref{eq30}).
 By comparing (\ref{eq30}) with the sum of asymptotic forms given by (\ref{eq48})-(\ref{eq52}),
 we can obtain the coefficient $A_k$. To this end, we need to expand $z$
 first,
 \be
 \label{eq53}
 xz=\sum\limits_{n=0}^{\infty}\frac{f_n}{n!}(\ln x)^n,
  \ee
  where $f_{n}/f_{0}$ is of order $n$ in the slow-roll
expansion. This expansion is useful for $\exp(-1/\xi)\ll x\ll
\exp(1/\xi)$. By using Eqs.~(\ref{eq10}), (\ref{eq26}),
(\ref{eq16}) and (\ref{eq19}), up to second-order, we obtain
 \be
 \label{eq54}
  f_{2}=\left.\frac{d^{2}(xz)}{(d\ln
x)^2}\right|_{x=1}\simeq \left.\frac{k}{H}
\sqrt{\frac{\epsilon_{1}}{2}}\left(16\epsilon_{1}^{2}
+9\epsilon_{1}\epsilon_{2}-\frac{1}{2}\epsilon_{2}^{2}
-\epsilon_{2}\epsilon_{3}\right)\right|_{aH=C_sk}, \ee \be
 \label{eq55}
f_{1}=\left.\frac{d(xz)}{d\ln x}\right|_{x=1}
 \simeq \left.\frac{k}{H}\sqrt{\frac{\epsilon_{1}}{2}}
 \left(-4\epsilon_{1}-\epsilon_{2}-12\epsilon_{1}^{2}
  -4c_{1}\epsilon_{1}^{2}-4\epsilon_{1}\epsilon_{2}
   -\frac{3}{2}c_{1}\epsilon_{1}\epsilon_{2}\right)\right|_{aH=C_sk},
\ee
\be
 \label{eq56}
  f_{0}\simeq  \left.\frac{k}{H}\sqrt{\frac{\epsilon_{1}}{2}}
  \left[2+(2+c_{1})\epsilon_{1}+\left(10+3c_{1}+\frac{3}{4}c_{1}^{2}
  +c_{2}\right)\epsilon_{1}^{2}+\left (3+\frac{1}{2} c_1 \right )\epsilon_{1}
  \epsilon_{2}\right]\right|_{aH=C_sk}.
\ee Then, up to second-order corrections, the asymptotic form
(\ref{eq30}) for $y(x)$ in the limit $x\to 0$ can be expressed as
 \be
\label{eq57}
 y(x)\to
\sqrt{2k}A_{k}f_{0}x^{-1}+\sqrt{2k}A_{k}f_{1}x^{-1}\ln
x+\frac{1}{2}\sqrt{2k}A_{k}f_{2}x^{-1}(\ln x)^{2}.
 \ee
 Collecting Eqs.~(\ref{eq48})-(\ref{eq52}) together also gives an asymptotic form
 for $y(x)$ up to second-order corrections. Note that
$$C_{sc}\simeq 1+{\cal O}(\xi)+{\cal O}(\xi^2),
~~~~~~ \frac{1}{C_{sc}}\simeq 1+{\cal O}(\xi)+{\cal O}(\xi^2),
~~~~~~\ln C_{sc}\simeq {\cal O}(\xi)+{\cal O}(\xi^2).$$
We can
therefore throw away the third and higher order corrections in
Eqs.~(\ref{eq48})-(\ref{eq52}) coming from $C_{sc}$.  Comparing
the result with Eq.~(\ref{eq57}), the coefficient of $x^{-1}$ will
give $A_k$ up to second-order corrections. The coefficients of
$x^{-1}\ln x$ and $x^{-1}(\ln x)^2$ simply give the consistent
asymptotic behavior, namely proportional to $z$. We finally arrive
at
\begin{eqnarray}
 \label{eq58}
A_k &=&\disp\frac{i}{\sqrt{2k}f_{0}}\left\{\frac{1}{C_{sc}}+\frac{c_{1}s_{2}}{4}
\left[-2-4i+3\alpha+\beta_{1}+i\beta_{2}-\frac{3i\pi}{2}\right]\right.
 \nonumber \\
 & & +\disp\frac{g_{1}}{3C_{sc}}\left[\alpha-\ln C_{sc}+\frac{i\pi}{2}\right]
 +\frac{g_{1}^{2}}{18}\left[\alpha^{2}-\frac{2}{3}\alpha-4+\frac{\pi^2}{4}
 +i\pi \left (\alpha-\frac{1}{3} \right )\right] \nonumber \\
& & \left. +\disp\frac{g_{2}}{6}\left[\alpha^{2}+\frac{2}{3}\alpha
  -\frac{\pi^2}{12}+i\pi \left (\alpha+\frac{1}{3} \right )\right]\right\},
\end{eqnarray}
where $\beta_{1}+i\beta_{2}\equiv 2ie^{2i}+3Ei(2i)=-0.5496523673+i\, 8.6963342377$.\\

 The power spectrum is defined by
 \be
  \label{eq59}
   {\cal P}_{s}(k)=\left(\frac{k^3}{2\pi^{2}}\right)\lim\limits_{x\to
0}\left|\frac{w_{k}}{z}\right|^{2}=\frac{k^3}{2\pi^{2}}|A_{k}|^{2}.
\ee Substituting Eqs.~(\ref{eq58}), (\ref{eq36}), (\ref{eq56}),
(\ref{eq28}), (\ref{eq29}), (\ref{eq33}) and (\ref{eq34}) into the
above formula, we obtain
\begin{eqnarray}
\label{eq60}
{\cal P}_{s}(k)&=&\disp\frac{H^2}{8\pi^{2}\epsilon_{1}}\left\{1+(4\alpha-2)\epsilon_{1}
 +\alpha\epsilon_{2}+\left[\, 4\alpha^{2}-23+\frac{7\pi^{2}}{3}+(2-\alpha-\beta_{1})c_{1}
 \right]\epsilon_{1}^{2}\right. \nonumber \\
& &
+\disp\left[\,\frac{3}{2}\alpha^{2}+\alpha-11+\frac{29\pi^{2}}{24}+\frac{1}{2}
(2-\alpha  -\beta_{1})c_{1}\right]\epsilon_{1}\epsilon_{2} \nonumber \\
& &\left. +\disp\left(\alpha^{2}-1+\frac{\pi^2}{12}\right)\epsilon_{2}^{2}+\frac{1}{2}
  \left(\alpha^{2}-\frac{\pi^2}{12}\right)\epsilon_{2}\epsilon_{3}\right\},
\end{eqnarray}
 where the right hand side should be evaluated at $aH=C_sk$. The
spectral index is defined by
 \be
  \label{eq61}
n_{s}(k)-1\equiv\frac{d\ln {\cal P}_{s}(k)}{d\ln k}.
 \ee
  It is easy to get the result in present case,
   \begin{eqnarray}
   \label{eq62}
 n_{s}(k)
 &=&  \disp 1-4\epsilon_{1}-\epsilon_{2}+8(\alpha-1)\epsilon_{1}^{2}
  -\alpha\epsilon_{2}^{2}+(5\alpha-3)\epsilon_{1}\epsilon_{2}-\alpha\epsilon_{2}\epsilon_{3}
    \nonumber \\
& &\disp
+4\left[-4\alpha^{2}+10\alpha-27+\frac{7\pi^2}{3}+(3-\alpha
 -\beta_{1})c_{1}\right]\epsilon_{1}^{3}+\left(2-\alpha^{2}-\frac{\pi^2}{6}\right)
  \epsilon_{2}^{3}  \nonumber \\
& &\disp
   +\frac{1}{2}\left[-31\alpha^{2}+60\alpha-172+\frac{199\pi^2}{12}+(20-7\alpha
   -7\beta_{1})c_{1}\right]\epsilon_{1}^{2}\epsilon_{2} \nonumber \\
& &\disp
  +\left(2-\frac{3}{2}\alpha^{2}-\frac{\pi^2}{8}\right)\epsilon_{2}^{2}\epsilon_{3}
   +\left(\alpha^{2}-\alpha-2+\frac{\pi^2}{6}+\frac{1}{2}c_1\right)\epsilon_{1}\epsilon_{2}^{2}
  \nonumber \\
  & &\disp
+\frac{1}{2}\left[7\alpha^{2}-8\alpha+22-\frac{31\pi^2}{12}+(\alpha+
    \beta_{1}-2)c_{1}\right]\epsilon_{1}\epsilon_{2}\epsilon_{3}
    +\frac{1}{2}\left(\alpha^{2}-\frac{\pi^2}{12}\right)\epsilon_{2}\epsilon_{3}\epsilon_{4},
\end{eqnarray}
where also the right hand side should be evaluated at $aH=C_sk$.


\sect{Some special cases}

\subsection{The case of canonical scalar field}

 In this case, $C_{s}^{2}=1$, namely $c_{1}=c_{2}=0$.
Substituting into Eqs.~(\ref{eq60}) and (\ref{eq62}), we will
obtain the power spectrum and spectral index for the case of
canonical scalar field. It is interesting to compare our results
with the one obtained in Ref.~\cite{s7}.  Having considered the
relations between the slow-roll parameters in Ref.~\cite{s7} and
in this paper given Eq.~(\ref{eq20}), and for the case of
canonical scalar field,
$$\epsilon_{1}=-\frac{\dot{H}}{H^2}=\frac{1}{2}\left(\frac{\dot{\phi}}{H}\right)^{2},$$
it is easy to see our results are completely the same as those of
Ref.~\cite{s7}. This gives a challenging check of our formalism.

\subsection{The case of tachyon inflation}

 In this case \cite{s11,s12}, $C_{s}^{2}=1-\dot{T}^{2}=1-\disp\frac{2}{3}\epsilon_{1}$,
  namely $c_{1}=\disp\frac{2}{3}$ and $c_{2}=0$. Substituting these into Eqs.~(\ref{eq60}) and
   (\ref{eq62}), we  obtain the power spectrum and spectral index for tachyon case.
     Note that the first order corrections have been calculated, for example,
     in \cite{s12}. The result of the second-order corrections to the power spectrum and
     spectral index for the case of tachyon inflation is {\it new}. This is an
    concrete example that the sound speed is time-dependant.  When one attempts to
     calculate the second  and higher order corrections, $C_{s}^{2}$ can not be
     taken as a constant approximately. Thus, the formalism presented in this paper
     is applicable in this case.

\subsection{The first-order corrections}

 When one considers only first-order corrections, the slow-roll parameters can be
 regarded approximately as some constants. From Eqs.~(\ref{eq60}) and (\ref{eq62}), for any
$C_{s}^{2}=1-c_{1}\epsilon_{1}-c_{2}\epsilon_{1}^{2}$, we have
 \be
 \label{eq63}
{\cal P}_{s}(k)\simeq\frac{H^2}{8\pi^{2}\epsilon_{1}}[1+
  (4\alpha-2)\epsilon_{1}+\alpha\epsilon_{2}],
\ee
 and
 \be
 \label{eq64}
  n_{s}(k)\simeq
   1-4\epsilon_{1}-\epsilon_{2}+8(\alpha-1)\epsilon_{1}^{2}-\alpha\epsilon_{2}^{2}
+(5\alpha-3)\epsilon_{1}\epsilon_{2}-\alpha\epsilon_{2}\epsilon_{3},
 \ee
  where again the right hand side should be evaluated at $aH=C_sk$.
  Note that in these expressions, the coefficients $c_1$ and $c_2$
  do not appear. This implies that up to the first order
  corrections, the time-dependence of sound speed does not have
  any effect on the power spectrum of density perturbation and spectral
  index.


In summary, we have extended Green's function method developed by
Stewart and Gong to calculate the power spectrum and its spectral
index of density fluctuation produced during inflation in the case
with a time-dependent sound speed, up to second order corrections
in the slow-roll expansion. The result for the tachyon inflation
is included as a special case. We have noted that up to the first
order corrections, there are no effects of the time-dependence of
sound speed on the power spectrum and spectral index.

\sect{Appendix}

 As mentioned at the end of section 2.2, we can deal with a class
of more general sound speed as
\be
C_{s}^{2}=1-c_{1}\epsilon_{m}-c_{2}\epsilon_{p}\epsilon_{q},
 \ee
where $m$, $p$ and $q$ are some positive integers. Similar to the
case of $\epsilon_{1}$, from Eqs.~(\ref{eq19}) and (\ref{eq26}),
we have
 \be \label{eq66}
 \frac{d\epsilon_{m}}{dx}\simeq -x^{-1}\epsilon_{m}\left[\epsilon_{1}
  +(-1)^{m+1}(\epsilon_{m}+\epsilon_{m+1})\right]
\ee for any $\epsilon_{m}$. And
$\disp\epsilon_{m}\left[\epsilon_{1}+(-1)^{m+1}(\epsilon_{m}+\epsilon_{m+1})\right]$
can be treated as a constant approximately because its derivative
with respect to time is a third-order small quantity which can be
ignored. Thus integrating Eq.~(\ref{eq66}) yields
 \be
  \label{eq67}
   \epsilon_{m}=s_{m1}+s_{m2}\ln x,
    \ee
 up to second-order, where
 \be
  \label{eq68}
s_{m1}\simeq\epsilon_{m}|_{aH=C_sk},\hspace{3mm} s_{m2}\simeq
   -\epsilon_{m}\left.\left[\epsilon_{1}+(-1)^{m+1}(\epsilon_{m}
   +\epsilon_{m+1})\right]\right|_{aH=C_sk}.
\ee
 We then can recast the sound speed as
 \be
  \label{eq69}
C_{s}^{2}=C_{sc}^{2}-c_{1}s_{m2}\ln x,
\ee
 up to second-order, where
  \be
  \label{eq70}
   C_{sc}^{2}=1-c_{1}s_{m1}-c_{2}s_{p1}s_{q1}.
   \ee
The formalism developed in section 3 is still valid for this
 $C_{s}^{2}$ because the calculations keep the forms of $C_{sc}$, $s_{n}$, $g_{n}$ and
 $f_{n}$ unchanged. The only difference is to replace the $s_2$ in $A_k$ by $s_{m2}$.
 Because the variable $z$ is dependent of $C_{s}$ as in Eq.~(\ref{eq10}), we must
 calculate the
 slow-roll expansions of $z$ and $\disp g(\ln x)=\frac{x^2}{k^2}\frac{z^{\prime\prime}}{z}-2$
 accordingly. Once having the new $f_0$, $g_1$ and $g_2$, and using the new
 $C_{sc}^{2}$ given in Eq.~(\ref{eq70}) and $s_{m1}$, $s_{m2}$ in Eq.~(\ref{eq68}),
 we then can get the corresponding $A_k$.  Thus the final power spectrum and spectral
index are in hand. The final results are very involved and their
expressions are long in length, we do not present them here.

\section*{Acknowledgments}
We thank Profs. Y.S. Myung and Y.Z. Zhang for useful discussions.
We are also grateful to Hong-Sheng Zhang, Qi Guo, Zong-Kuan Guo,
Da-Wei Pang, Xun Su, Ren Wei, and Ding-Fang Zeng for discussions
and help.
 One of the authors (RGC) would like to express his
gratitude to the Physics Department, Baylor University for its
hospitality. This work was supported by Baylor University, a grant
from Chinese Academy of Sciences, a grant from NSFC, China (No.
13325525), and a grant from the Ministry of Science and Technology
of China (No. TG1999075401).


\begin{thebibliography}{99}
\bibitem{s1} A.H. Guth, Phys. Rev. D23 (1981) 347;\\
           A.D. Linde, Phys. Lett. B108 (1982) 389;\\
           A. Albrecht and P.J. Steinhardt, Phys. Rev. Lett. 48 (1982) 1220;\\
           A.D. Linde, {\tt hep-th/0402051}.
\bibitem{s2} For example, H.V. Peiris et al., Astrophys. J. Suppl. 148 (2003) 213 {\tt [astro-ph/0302225]};\\
           A.R. Liddle and D.H. Lyth, "Cosmological inflation and large-scale structure", Cambridge Univ. press, 2000.
\bibitem{s3} For example, C.L. Bennett et al., Astrophys. J. 464 (1996) L1 {\tt [astro-ph/9601067]};\\
           A.H. Jaffe et al., Phys. Rev. Lett. 86 (2001) 3475 {\tt [astro-ph/0007333]};\\
           N.W. Halverson et al., Astrophys. J. 568 (2002) 38 {\tt [astro-ph/0104489]};\\
           P.P. van der Werf et al., {\tt astro-ph/0010459};\\
           C. Pryke et al., Astrophys. J. 568 (2002) 46 {\tt
           [astro-ph/0104490]}.
\bibitem{s4} C.L. Bennett et al., Astrophys. J. Suppl. 148 (2003) 1 [{\tt astro-ph/0302207}];\\
           D.N.Spergel et al., Astrophys. J. Suppl. 148 (2003) 175 [{\tt astro-ph/0302209}].\\
           The official website of WMAP is {\tt
           http://map.gsfc.nasa.gov}\\
           The data of WMAP are released on {\tt
           http://lambda.gsfc.nasa.gov}.
\bibitem{s5} M. Tegmark et al., Phys. Rev. D69 (2004) 103501 {\tt
[astro-ph/0310723]};\\
            SDSS Collaboration, astro-ph/0410239; astro-ph/0403325; astro-ph/0305492; \\
           The official website of SDSS is {\tt
           http://www.sdss.org}.
\bibitem{s6} For example, A.R. Liddle and D.H. Lyth, Phys. Rept. 231 (1993) 1 {\tt [astro-ph/9303019]};\\
           V.F. Mukhanov, H.A. Feldman and R.H.Brandenberger, Phys. Rept. 215 (1992) 203;\\
           J.M.Bardeen, Phys. Rev. D22 (1980) 1882.

\bibitem{insert1}E.~D.~Stewart and D.~H.~Lyth,
Phys.\ Lett.\ B {\bf 302}, 171 (1993) [arXiv:gr-qc/9302019].

\bibitem{s7} E.D. Stewart and J.-O. Gong, Phys. Lett. B510 (2001) 1.
\bibitem{s8} K.H. Kim, H.W. Lee and Y.S. Myung, Phys. Rev. D70 (2004) 027302;\\
           H.-S. Kim, G.S. Lee, H.W. Lee and Y.S. Myung, Phys. Rev.
           D70 (2004) 043521.
\bibitem{s9} J. Garriga and V.F. Mukhanov, Phys. Lett.
            B458 (1999) 219.
\bibitem{s10} C. Armend${\rm \acute{a}}$riz-Pic${\rm \acute{o}}$n,
      T. Damour and V. Mukhanov, Phys. Lett. B458 (1999) 209.
\bibitem{s11} G.W.Gibbons, Phys. Lett. B537 (2002) 1 {\tt [hep-th/0204008]};\\
            M. Fairbairn and M.H.G. Tytgat, Phys. Lett. B546 (2002) 1 {\tt [hep-th/0204070]};\\
            A. Frolov, L. Kofman and A. Starobinsky, Phys. Lett. B545 (2002) 8 {\tt [hep-th/0204187]};\\
            Y.-S. Piao, R.-G. Cai, X.-M. Zhang and Y.-Z. Zhang, Phys. Rev.
             D66 (2002)121301 {\tt [hep-ph/0207143]}; \\
      G.~W.~Gibbons,
   Class.\ Quant.\ Grav.\  {\bf 20}, S321 (2003)
[arXiv:hep-th/0301117];

D.~A.~Steer and F.~Vernizzi,
Phys.\ Rev.\ D {\bf 70}, 043527 (2004) [arXiv:hep-th/0310139];

G.~Calcagni,
Phys.\ Rev.\ D {\bf 69}, 103508 (2004) [arXiv:hep-ph/0402126].

\bibitem{s12} G. Calcagni, {\tt hep-ph/0406057}.

\end{thebibliography}
\end{document}